\documentclass [twocolumn,showpacs,prl]{revtex4}
\usepackage{graphicx}
\usepackage{bm}
\narrowtext
\begin{document}
\title{Role of surface roughness in hard x-ray emission from femtosecond laser produced copper plasmas}   
\author{P. P. Rajeev, S. Banerjee\footnote{Present address: CUOS, University of Michigan, Ann Arbor, 
MI, U. S. A.}, A. S. Sandhu, R. C. Issac\footnote{Present address: Dept. of Physics, University of Strathclyde, Scotland, U. K.},
 L. C. Tribedi and G. R. Kumar\footnote{electronic mail: grk@tifr.res.in}}
\address{Tata Institute of Fundamental Research, Homi Bhabha 
Road, Mumbai 400 005, India.}
\date{\today}

\begin{abstract}
The hard x-ray emission in the energy range of $30-300$ keV from copper plasmas produced by 100 fs, 806 nm laser pulses  at intensities in the range of 10$^{15}$-10$^{16}$ W cm$^{-2}$ is investigated. We demonstrate that surface roughness of the targets overrides the role of polarization state in the coupling of light to the plasma. We further show that surface roughness has a significant role in enhancing the x-ray emission in the above mentioned energy range.     
\end{abstract}

\maketitle

	The behaviour of matter under extremely intense, ultrashort light pulse irradiation is an exciting area of contemporary research \cite{Gibbon}. Highly dense plasmas with steep density gradients and temperatures of hundreds of electron volts can be produced at the focal spot of an intense, femtosecond laser. Such plasmas are remarkably different from conventional laboratory plasmas as they are formed rapidly and hydrodynamic expansion is insignificant during the laser pulse. They are `point' (micron size) sources for both soft and hard x-ray \cite{Kieffer,Chen,Giuleitti} and gamma ray pulses \cite{Kmetec,Zhang}. This aspect has attracted multifaceted research to explore various applications like x-ray lithography and time resolved x-ray diffraction \cite{Gibbon}. In addition to the large yield both in continuum and line emissions, an exciting property of such x-ray pulses \cite{Murnane} is their extremely short temporal duration  (subpicosecond), which is ideal for time resolved studies at x-ray wavelengths. To be able to use such x-ray sources, it is essential to simply and correctly characterize their emission as well as to find ways of enhancing it. Recently, Banerjee et al. have demonstrated a simple way of obtaining absolute yields of such x-ray fluxes and have pointed out the role of photon statistics in estimating yields from laser produced plasmas using broad band Si (Li) detectors \cite{SB1,SB2}. There is a great deal of interest in methods that could enhance the x-ray yield, and the influence of various laser and target conditions has been the subject of many recent studies. Pre-plasma formation has been investigated in detail as one of the prominent ways of improving the x-ray yields. While significant enhancement in the emission is noticed, the x-ray pulse duration tends to become longer in such cases \cite{Pelletier,Nakano}.  There is increasing interest on the role of modulation/roughness of the surface in increasing the coupling of the input light into the plasma, which results in an enhancement in the x-ray yield. Murnane and coworkers \cite{Falcone} have shown absorption of over $90\%$ input light into the plasma formed on grating targets as well as those coated with metal clusters. More recently, impressive enhancements of x-ray flux have been achieved in nanohole alumina targets (soft x-ray region) \cite{Nishikawa1}, porous silicon \cite{Nishikawa2} and nickel `velvet' targets (hard x-ray region) \cite{Kulcsar}. There have, however, been no reports of enhancements in the very hard region ($\ge$10 keV).  Such studies should be interesting not only from the point of view of enhanced hard x-ray emission, but also to understand the role of surface structure in the generation of hot electrons which are responsible for the emission. During the course of our studies of bremsstrahlung emission in the hard and very hard x-ray regimes, we observed that unpolished targets showed a significant enhancement in the x-ray yield as compared to polished ones. In this paper, we present measurements of the bremsstrahlung emission in the $30-300$ keV region from polished and unpolished copper targets and emphasize the influence of roughness on the yield and polarization dependence of the x-ray emission.

	 A Ti:Sapphire  laser (806 nm, 100 fs) was focussed with a 30 cm focal length lens on   copper targets housed in a vacuum chamber at 10$^{-3}$ torr. The femtosecond laser is a custom-built chirped pulse amplification system with two-stages of multipass amplification \cite{SB2}. The maximum pulse energy used in the current experiments is 6 mJ, giving a peak intensity of about $2{\times}10^{16}$ W cm$^{-2}$ at the focal point of diameter 30 $\mu$.  The laser had a prepulse  (13 ns ahead of the main pulse) that was at least 10$^4$ times weaker and the contrast with the pedestal(at 100 ps) was better than 10$^5$. Under these conditions, plasma formation by prepulse/pedestal is found to be negligible \cite{OCzhang}. A thin half wave plate was introduced in the beam path in order to change the polarization states. The target was constantly rotated and translated in order to avoid multiple hits at the same spot by the laser pulses. X-ray emission from the plasma was measured along the plume direction by a NaI(Tl) detector. The detector was shielded by lead bricks and calibrated using Co$^{57}$, Cs$^{134}$ and Eu$^{152}$. The BK-7 window of the vacuum chamber sets a low energy cutoff at about 12 keV for the observed emission. The signal from the detector was amplified and then fed to a multichannel analyzer through an analog to digital converter (ADC). Spectra were typically collected over $30000-40000$ laser shots. To ensure their reliability, the temperature fits presented are done using the data above 50 keV, wherefrom the transmission is $100\%$. In order to minimize the probability for pile-up, the count rate was reduced to less than 0.1 per laser shot by introducing suitable lead apertures in front of the detector.  The spectra were measured at different distances from the target (different solid angles) and the detector was finally placed at a position where there was no pile up. Typical solid angle of observation at this position was in the region of $50-80 \mu$sr.The spectra were made nearly background free by eliminating cosmic ray noise by means of time gating - the laser pulse trigger was sent to a delay gate generator which activated a time window of 10-20 microseconds for the signal acquisition.  

\begin{figure}
\centering
\includegraphics [width=2.8in,height=3in]{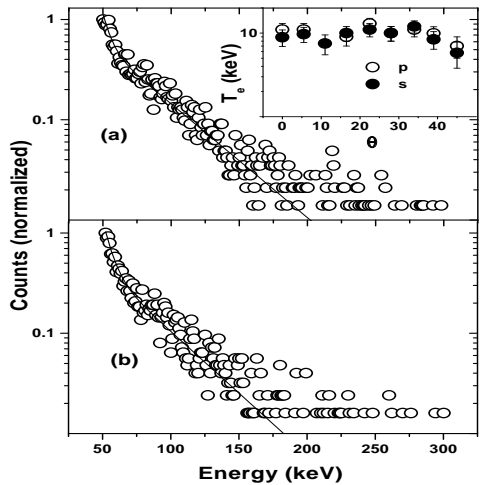}
\caption{Bremsstrahlung emission from unpolished copper target irradiated at $1.6{\times}10^{16}$ W cm$^{-2}$ with (a) p - polarization and (b) s - polarization. Data points obatined over 36000 shots in the range $50 - 300$ keV are shown wherein the exponential fits are performed. The inset in (a) shows the temperature as a function of laser incident angle for both polarizations.}
\end{figure}

	Fig. 1 shows the bremsstrahlung emission measured in the $30-300$ keV region for an  unpolished copper target at an intensity of $1.6{\times}10^{16}$ W cm$^{-2}$, with p and s - polarized light incident at 45$^{\circ}$.  The solid line fits indicate the existence of at least two temperature components for the hot electrons in the plasma. These turn out to be 6 $\pm$ 1 keV and 35 $\pm$ 5 keV in both cases, irrespective of the polarization state of the light field. The space and energy integrated yield, under the assumption of isotropic emission, gives an overall efficiency (keV/keV) of about $2{\times}10^{-3}$ for conversion into the $30-300$ keV region. 

	 In the case of a p-polarized laser beam, it is well known that in experimental conditions similar to ours, the hot electrons could be generated by two mechanisms - resonance absorption (RA) and vacuum heating (VH). RA has been well studied both experimentally \cite{Schnurer,Bastiani,Teubner1,Teubner2} and theoretically \cite{Forslund,Estabrook,Milchberg,Fedosejev} and based on the observations and simulations, the following scaling law \cite{Forslund} has been established:

\begin{equation}
T_{hot} = 14{\times}T_c^{0.33}(I\lambda^{2})^{0.33}
\end{equation}

where $T_c$ is the background electron temperature in keV, I is the intensity of the laser in units of 10$^{16}$ W cm$^{-2}$ and $\lambda$ is the wavelength in microns. According to this scaling law, for a $T_c$ of 0.1 keV, we get a $T_{hot}$ of 6.6 keV under our experimental conditions. This temperature is close to the lower component that we 
have measured in our experiments. RA cannot, however, explain the higher component.  VH \cite{Brunel,Grimes} can be examined as a possible candidate for the generation of this component. A crucial requirement for VH is that the electron oscillatory amplitude ($x_{osc}$) is larger than the plasma scale length ($L$). Our laser pulses have an insignificant pre-pulse component and the measurements of Doppler shift from plasma expansion have indicated \cite{AS}
 that the ratio $\frac{\rm x_{osc}}{\rm L}$ is about 1 and $\frac{\rm L}{\rm \lambda}$ is about 0.01. For these parameters, simulations \cite{Gibbonbell} show that VH could give a $T_{hot}$ of about 30 keV, under similar conditions, which is quite close to the measured value of the higher temperature component (reflectivity studies are done which support the existence of VH under our conditions). 

\begin{figure}
\centering
\includegraphics [width=2.8in,height=3in]{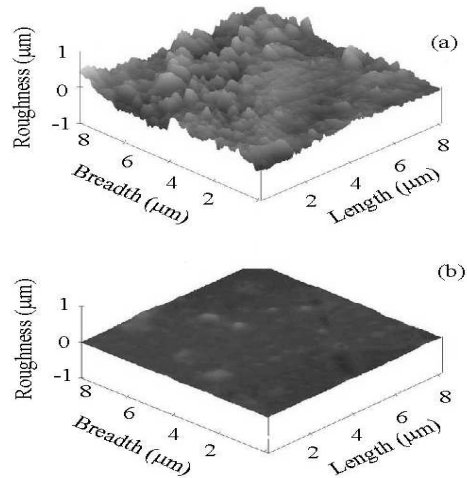}
\caption{AFM images of the targets used. (a) unpolished copper with average roughness of $0.1 \mu$ ,(b) polished copper with average roughness of $5 nm$.}
\end{figure}

	Such an interpretation, however, has to be examined in the light of the data for s-polarization. In this case, there is no known mechanism (operative under our experimental conditions) that can explain the lower hot electron temperature, let alone the higher one.  The only mechanism that one can invoke is collisional absorption, which becomes less effective \cite{Wulker} above 10$^{15}$ W cm$^{-2}$. Surprisingly, not only are there high temperature components in our s-polarization data, but their magnitudes are also quite comparable to those generated by p-polarized light. The total (integrated) energy of x-rays emitted in the range $30-300$ keV using p-polarization is calculated to be $3.5{\times}10^{5}$ keV where as it is $2.8{\times}10^{5}$ keV using s-polarized light, again contrary to expectation. The inset in Fig. 1(a) shows temperature, measured at a lower intensity of 2 x 10$^{15}$ W cm$^{-2}$. It is nearly constant with laser incidence angle for both polarizations. We noted similar behaviour at other intensities also.

	These observations demand an examination of the possible role of surface roughness of targets in the hot electron generation in our experiments. It has been pointed out in many studies that a number of efficient schemes - surface waves, multiple scattering, trapping of energetic electrons and light etc. exist for the coupling of laser light into the plasma for rough surfaces. The understanding thus far is that structuring of the surface leads to localized volume heating of micro regions of the target leading to denser plasmas and higher temperatures \cite{Falcone,Nishikawa1,Nishikawa2,Kulcsar}. 
To investigate the level of roughness on our surface,  an Atomic Force Microscope (AFM) image of the target, shown in Fig. 2(a) is taken. It is evident that the surface is quite uneven, with the average peak-valley difference being $0.1 \mu$. To understand the role of roughness clearly, the x-ray emission from a highly polished surface is investigated. The AFM image of this target is shown in Fig.2 (b). This surface has an average peak-valley separation of 5 nm. This polished target is clearly 20 times smoother than the earlier unpolished target that we discussed. 

\begin{figure}
\centering
\includegraphics [width=2.8in,height=3in]{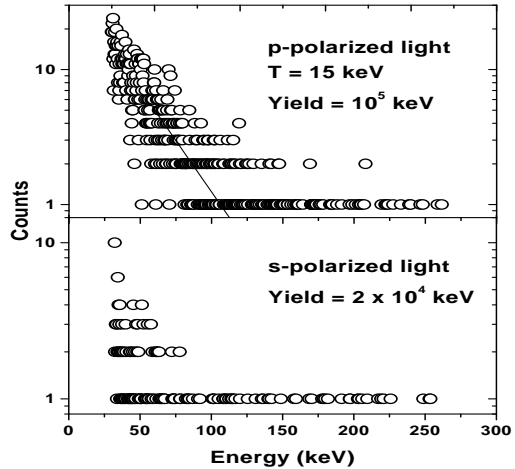}
\caption{Bremsstrahlung emission over 36000 laser shots from a polished copper target irradiated at $9{\times}10^{15}$ W cm$^{-2}$. Temperature is not obtained for the data for s-polarization because of inadequate counts.}
\end{figure}

	Fig. 3 shows the bremsstrahlung data from polished surface irradiated at $9{\times}10^{15}$ W cm$^{-2}$ under other conditions similar to those described above for unpolished targets. The differences in the spectra are striking. The yield obtained (in the range $30-300$ keV) using p-polarization is about 5 times larger than that obtained using s-polarization as expected by the large coupling of the former into the plasma by RA and VH. The exponential fit for the p-polarization data gives a temperature of 15 $\pm $ 3 keV . This temperature component is again on par with the value obtained from simulations for the parameters valid at this intensity \cite{Gibbonbell}. No fit is attempted on the data obtained using s-polarized light as the counts beyond 50 keV are too few to get a good fit. We obtained similar data at other intensities also and the general features remain the same.
	
	Fig. 4(a) presents a comparative picture of bremsstrahlung emission from unpolished and polished targets irradiated with p-polarization at $1.4{\times}10^{16}$ W cm$^{-2}$ at $45 ^\circ $ . There is an enhancement of around $60\%$ in the total x-ray energy emitted (in the range $30-300$ keV) from the unpolished target as compared to the polished one.   Fig 4(b) shows the variation of x-ray emission with angle of incidence for rough and smooth targets keeping the intensity (corrected for oblique incidence) constant at 1.2$\times$10$^{16}$ W cm$^{-2}$, using p-polarized light throughout. It demonstrates that rough targets give enhanced yields at all angles. Inset in Fig. 4(a) are the x-ray spectra taken at the best angles for rough and smooth targets, which shows that the best yield from the smooth target is  one-fourth of the best yield of a rough target. These observations lead us to infer that the roughness on the unpolished target is responsible for the increase in the bremsstrahlung yield.  However, the temperatures obtained are nearly the same (22 $\pm$ 4 keV) for both polished and unpolished targets at $45 ^\circ $.

\begin{figure}
\centering
\includegraphics [width=2.8in,height=3in]{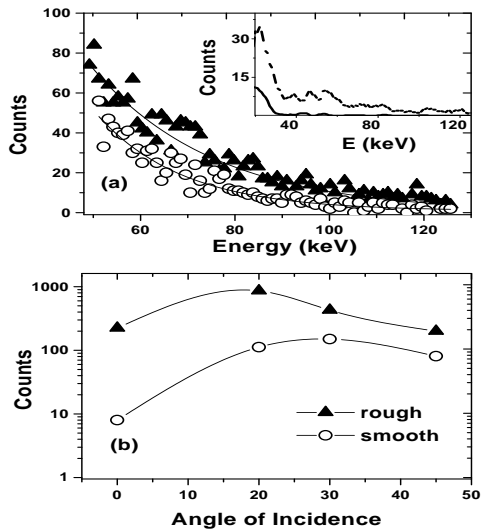}
\caption{(a): Comparison of bremsstrahlung emission from  smooth (open circles) and rough (solid triangles) targets irradiated with p-polarized light at $45 ^\circ $. Data points obatined over 36000 shots in the range $50 - 125$ keV are shown wherein the exponential fits are performed. (b) Spectrally integrated emission, collected over 2000 laser shots, from smooth and rough targets as a function of angle of incidence. The inset of (a) is the comparison of bremsstrahlung emission from smooth (solid line) and rough (dashed line) targets at their respective angular maxima shown in (b).}
\end{figure}

	It is noted that a rough target gives more x-ray yield and higher temperatures compared to its smooth counterpart when irradiated by s-polarized light. The local topographical features present in the case of rough target can, in principle, make the laser incident at different angles locally. Because of the lack of a definite geometry on the surface, the incident s-polarization could be considered as `p' at some points, depending on the local target morphology, which could eventually give rise to hot electrons due to RA and VH.
However, this effect would have a detrimental impact for p-polarized light incident on rough targets since the polarization can be locally `s', again depending on the local geometry, which would mean less amount of x-ray emission as compared to that from polished targets. Since the yields are invariably higher in the case of a rough target even with a p-polarized light field, irrespective of the angle of incidence, other mechanisms that would enhance the laser-plasma coupling, viz. surface waves, electron confinement may need to be invoked. A detailed study is necessary in this regard and is under way. 
 
	Another interesting feature that emerges, is the relative lack of dependence of the x-ray emission on the polarization in case of the unpolished target. This could mainly be attributed to two factors - (i). change of polarization due to local morphology, and (ii). modification of polarization by scattering from the rough features. These factors imply that the influence of roughness could nullify the differences in light coupling to the plasma in the case of s and p-polarized light fields. We note that Ahn {\it{et al.}} \cite{Ahn}, have seen similar lack of dependence of soft x-ray yield on the light polarization state and cited rippling of the critical surface as a possible cause. More investigations are needed to study this problem in detail and we are in the process of studying the influence of controlled modification of surface 
roughness on hard and very hard x-ray emission. 

	Rippling of critical surface (as an analog of roughness) has been  examined theoretically in some earlier studies as a possible cause of excess absorption and its partial independence of polarization\cite{Cairns,Thomson}. We must emphasize that the experimental evidence for such a possibility \cite{Manes,Maaswinkel,Schnurer96}, however, has been very tentative. These experimental results pointed to rippling as one of the possible reasons for deviations of measured absorption values from their agreement with those expected from established models of absorption like RA. Besides, they did not present any data on hot electron temperatures at all. We believe that our study is the first to qualitatively establish correlation of temperature with structural roughness. The advantage of our study is that we deal with measurable and predefined roughness in our targets.

	In conclusion, we have demonstrated that the roughness of ordinary, unpolished, readily available surfaces could be used to produce enhanced yields of hot electrons which in turn lead to larger fluxes of ultrashort x-ray pulses in the very hard x-ray region.  A key observation is the lack of influence of the polarization state on the hot electron temperatures and yields in the case of unpolished targets in stark contrast to the observations for polished targets. Further experiments are under way to study the hot electron generation in targets with tailored roughness.

	We thank A. Dharmadhikari for help in experiments, S. P. Pai for the AFM images, P. K. Kaw and S. Sengupta for discussions. The high energy, femtosecond laser facility has received substantial funding from the Department of Science and Technology, Government of India, New Delhi.






\begin{references}
\bibitem{Gibbon} P. Gibbon, and E. Forster, Plasma Phys. and Control. Fusion $\bf{38}$, 769 (1996)
\bibitem{Kieffer}	J. C. Kieffer,{\it{et al.}}, Appl. Optics $\bf{32}$, 4247 (1993)
\bibitem{Chen}	H. Chen, B. Soom, B. Yaakobi, S. Uchida, and D. D. Meyerhofer, Phys. Rev. Lett. $\bf{70}$, 3431 (1993)
\bibitem{Giuleitti} D. Giulietti, and L. A. Gizzi, La Rivista del Nuovo Cimento $\bf{21}$, 1 (1998)
\bibitem{Kmetec}	J. D. Kmetec, IEEE J. Quant. Electr. $\bf{28}$, 2382(1992)
\bibitem{Zhang}	P. Zhang {\it{et al.}}, Phys. Rev. E $\bf{57}$, R3746 (1998)
\bibitem{Murnane}	M. Murnane, H. Kapteyn, M. D. Rosen, and R. Falcone, Science $\bf{251}$, 531(1991)
\bibitem{SB1}	S. Banerjee, G. R. Kumar, A. K. Saha, and L. C. Tribedi, Optics Communications $\bf{158}$, 72 (1998); 
			Phys. Scr. $\bf{T80}$, 539 (1999); S. Banerjee, G. R. Kumar, and L. C. Tribedi in K. K. Sud, 
			and U. N. Upadhyaya (Eds) {\it{Trends in Atomic and Molecular Physics}} (Kluwer, 2000), p1.
\bibitem{SB2}	S. Banerjee, G. R. Kumar, and L. C. Tribedi, Eur. Phys. J. D $\bf{11}$, 295 (2000)
\bibitem{OCzhang} Y. Zhang, J.Zhang, S-h Pan and Y-x. Nie, Opt.Commun {\bf 126},85 (1996)
\bibitem{Pelletier}	J.F. Pelletier, M. Chaker and J.-C. Kieffer, J. Appl. Phys. $\bf{81}$, 5980 (1997) 
\bibitem{Nakano}	H. Nakano, T. Nishikawa, H. Ahn, and N. Uesugi, Appl. Phys. Lett. $\bf{69}$, 2992(1996)
\bibitem{Falcone}	M. M. Murnane, H. C. Kapteyn, S. P. Gordon, J. Bokor, E. N. Glytsis, and R. W. Falcone, Appl. Phys. Lett. $\bf{62}$, 1068(1993)
\bibitem{Nishikawa1}	Tadashi Nishikawa, Hidetoshi Nakano, Naoshi Uesugi, Masashi Nakao, and Hideki Masuda, Appl. Phys. Lett. $\bf{75}$, 4079(1999)
\bibitem{Nishikawa2}	Tadashi Nishikawa, Hidetoshi Nakano, Hyeyoung Ahn, and Naoshi Uesugi, Appl. Phys. Lett. $\bf{70}$, 1653(1997)
\bibitem{Kulcsar}	G. Kulcsar, D. AlMawlawi, F. W. Budnik, P.R. Herman, M. Moskovits, L. Zhao, and R. S. Marjoribanks, Phys. Rev. Lett. $\bf{84}$, 5149(2000)
\bibitem{Schnurer} M. Schnurer, R. Nolte, T. Schlegel, M. P. Kalachnikov, P. V. Nickels, P. Ambrosi, and W. Sandner, J. Phys. B: At. Mol. Opt. Phys. $\bf{30}$, 4653(1997)
\bibitem{Bastiani} S. Bastiani, P. Audebert, J. P. Geindre, T. Schlegel, J. C. Gauthier, C. Quoix, G. Hamoniaux, G. Grillon, and A. Antonetti, Phys. Rev. E $\bf{60}$, 3439(1999)
\bibitem{Teubner1}R. Sauerbrey, J.Fure, S.P. Le Blanc, B. van Wonterghem, U.Teubner and F.P. Schafer, Phys.Plasmas, {\bf 1}, 1635 (1994)
\bibitem{Teubner2} U. Teubner, I. Uschmann, P. Gibbon, D. Altenbernd, E. Förster, T. Feurer, W. Theobald, R. Sauerbrey, G. Hirst, M. H. Key, J. Lister, and D. Neely, Phys. Rev. E, {\bf 54}, 4167 (1996)

\bibitem{Forslund}D. W. Forslund, J. M. Kindel, and K. Lee, Phys. Rev. Lett. $\bf{39}$, 284(1977)
\bibitem{Estabrook}	K. Estabrook and W. L. Kruer, Phys. Rev. Lett. ${\bf 40}$, 42 (1978) 
\bibitem{Milchberg} H.M. Milchberg and R.R. Freeman, J.Opt. Soc. Am. B {\bf 6}, 1351 (1989) 
\bibitem{Fedosejev}R. Fedosejevs, R.Ottmann, R.Sigel, G. Kuhnle, S. Szatmari, and F.P.Schafer, Appl. Phys.B {\bf 50} 79(1990)

\bibitem{Brunel}	F. Brunel, Phys. Rev. lett. $\bf{59}$, 52 (1987)
\bibitem{Grimes}	M. K. Grimes, A. R. Rundquist, Y. S. Lee, and M. C. Downer, Phys. Rev. Lett. $\bf{82}$, 4010 (1999)
\bibitem{AS}	A. S. Sandhu( private communication)
\bibitem{Gibbonbell}	Paul Gibbon, and A. R. Bell, Phys. Rev. Lett. $\bf{68}$, 1535 (1992)
\bibitem{Wulker}	C. Wulker, W. Theobald, D. R. Gnass, F. P. Schafer, J. S. Bakos, R. Sauerbrey, S. P. Gordon, and R. W. Falcone, Appl. Phys. Lett. $\bf{68}$, 1338 (1996)
\bibitem{Silin}	V.P. Silin, Sov. Phys.JETP, $\bf{20}$, 1510 (1965); R.J. Faehl and N.F. Roderick, Phys. Fluids $\bf{21}$, 793 (1978)
\bibitem{Ahn}	H. Ahn, H. Nakano, T. Nishikawa, and N. Uesugi, Jpn. J. Appl. Phys, (Pt.2 Lett). $\bf{35}$, L154 (1996)
\bibitem{Cairns} R.A. Cairns, Plasma Phys. {\bf 20} 991 (1978)
\bibitem{Thomson} J.J. Thomson, W.L. Kruer, A.B. Langdon,C.E. Max and W.C. Mead, Phys. Fluids {\bf 21},707 (1978)
\bibitem{Manes} K.R.Manes, V.C.Rupert, J.M. Auerbach, P.Lee and J.E.Swain, Phys.Rev.Let. {\bf 39} 281 (1977)
\bibitem {Maaswinkel} A.G.M. Maaswinkel, K.Eidmann and R.Sigel Phys.Rev.Lett. {\bf 42},1625 (1979)
\bibitem{Schnurer96} M.Schnurer, P.V.Nickles, M.P.Kalachnikov, W. Sandner, R.Nolte, P.Ambrosi, J.L.Miquel, A.Dulieu and A.Jolas, J.Appl.Phys. {\bf 80} 5604 (1996)

\end{references}
\end{document}